\documentclass[aps,prl,superscriptaddress,twocolumn]{revtex4-1}
\usepackage{graphicx}
\usepackage[dvipsnames]{xcolor}
\usepackage{siunitx}
\usepackage{hyperref}
\usepackage{xfrac}
\usepackage{soul}
\usepackage{float}
\usepackage{color}
\usepackage{stmaryrd}
\usepackage{mathtools}
\usepackage{chemformula}
\usepackage[nameinlink,capitalise]{cleveref}
\hypersetup{
colorlinks=true,
linkcolor=blue,
filecolor=magenta,
urlcolor=cyan,
}

\makeatletter
\AtBeginDocument{\let\LS@rot\@undefined}
\makeatother

\begin{document}
\title{Optical Transmission Enhancement of Ionic Crystals via Superionic Fluoride Transfer: Growing VUV-Transparent Radioactive Crystals}
\author{Kjeld Beeks}
\author{Tomas Sikorsky}
\author{Fabian Schaden}
\author{Martin Pressler}
\author{Felix Schneider}
\affiliation{Institute for Atomic and Subatomic Physics, TU Wien, Stadionallee 2, 1020, Vienna, Austria}
\author{Björn N. Koch}
\affiliation{Anorganische Chemie, Fluorchemie, Fachbereich Chemie, Phillips-Universität Marburg, Hans-Meerwein-Str. 4, 35032 Marburg}
\author{Thomas Pronebner}
\author{David Werban}
\author{Niyusha Hosseini}
\author{Georgy Kazakov}
\affiliation{Institute for Atomic and Subatomic Physics, TU Wien, Stadionallee 2, 1020, Vienna, Austria}
\author{Jan Welch}
\author{Johannes H. Sterba}
\affiliation{CLIP, TRIGA Center Atominstitut, TU Wien, Stadionallee 2, 1020, Vienna, Austria}
\author{Florian Kraus}
\affiliation{Anorganische Chemie, Fluorchemie, Fachbereich Chemie, Phillips-Universität Marburg, Hans-Meerwein-Str. 4, 35032 Marburg}
\author{Thorsten Schumm}
\affiliation{Institute for Atomic and Subatomic Physics, TU Wien, Stadionallee 2, 1020, Vienna, Austria}
\date{\today}
\begin{abstract}
The \SI{8}{\electronvolt} first nuclear excited state in \ch{^{229}Th} is a candidate for implementing a nuclear clock. Doping \ch{^{229}Th} into ionic crystals such as \ch{CaF2} is expected to suppress non-radiative decay, enabling nuclear spectroscopy and the realization of a solid-state optical clock. Yet, the inherent radioactivity of \ch{^{229}Th} prohibits the growth of high-quality single crystals with high \ch{^{229}Th} concentration; radiolysis causes fluoride loss, increasing absorption at \SI{8}{\electronvolt}. These radioactively doped crystals are thus a unique material for which a deeper analysis of the physical effects of radioactivity on growth, crystal structure and electronic properties is presented. Following the analysis, we overcome the increase in absorption at \SI{8}{\electronvolt} by annealing \ch{^{229}Th} doped \ch{CaF2} at \SI{1250}{\degreeCelsius} in \ch{CF4}. This technique allows to adjust the fluoride content without crystal melting, preserving its single-crystal structure. Superionic state annealing ensures rapid fluoride distribution, creating fully transparent and radiation-hard crystals. This approach enables control over the charge state of dopants which can be used in deep UV optics, laser crystals, scintillators, and nuclear clocks.
\end{abstract}

\maketitle

Thorium-229 (\ch{^{229}Th}), possessing an \SI{8}{\electronvolt} and approximately 600-second lifetime first nuclear excited state (isomer state), enables high-precision vacuum ultraviolet (VUV) laser spectroscopy~\cite{Seiferle2019,kraemer2023observation,Masuda2019}. The anomalously low energy of this excited state offers the potential for the construction of an optical clock based on a nuclear transition~\cite{Peik2003,beeks2021thorium}. The structure of the nuclear levels is governed by both Coulomb and nuclear forces~\cite{hayes2007sensitivity}. This allows probing of these forces via nuclear spectroscopy, paving the way for new fundamental research: For example the search for dark matter, or potential drifts in the fine-structure constant~\cite{Peik_2021,Pavel2020sensitivity}.

\ch{^{229}Th} is required to be in a 3+ or higher charge state to suppress the non-radiative decay~\cite{Peik2015}. The possibility of trapping charged \ch{^{229}Th} in a solid-state matrix offers an alternative to ion traps. In the solid, the nucleus is isolated due to the small interaction with its chemical environment~\cite{Kazakov_2012}. Ionic crystals such as \ch{CaF2} are an excellent choice as host material for the \ch{^{229}Th} based nuclear clock~\cite{HEHLEN201391}. The ionic character of these crystals naturally forces the \ch{^{229}Th} into a 4+ charged state, substituting the calcium (\ch{Ca^2+}) cation~\cite{pimon2022ab,gong2024structures}, and their large band gaps make them transparent to wavelengths around \SI{150}{\nano\meter} or \SI{8}{\electronvolt}~\cite{rubloff1972far}. Ionic crystals such as oxides and fluorides display good scintillator properties and are resistant to VUV radiation, making them suitable to host the radioactive \ch{^{229}Th} and observe its radiative decay~\cite{TakayukiYANAGIDA2018PJA9402B-02,kraemer2023observation}. In this experiment, calcium fluoride (\ch{CaF2}) was chosen as the host material due to its excellent scintillator properties~\cite{rodnyi1997physical}, simple cubic structure~\cite{Villars2016:sm_isp_sd_0378096}, large \SI{12}{\eV} electronic and \SI{10}{\eV} optical bandgap~\cite{rubloff1972far}, and an unchanged \SI{10}{\eV} optical bandgap after thorium~\cite{beeks2022nuclear} doping. Both \ch{^{232}Th} and \ch{^{229}Th} were used to grow single crystalline Th:\ch{CaF2} using the vertical gradient freeze method~\cite{beeks2023growth}. 

\begin{figure*}[]
    \includegraphics[width=0.8\textwidth]{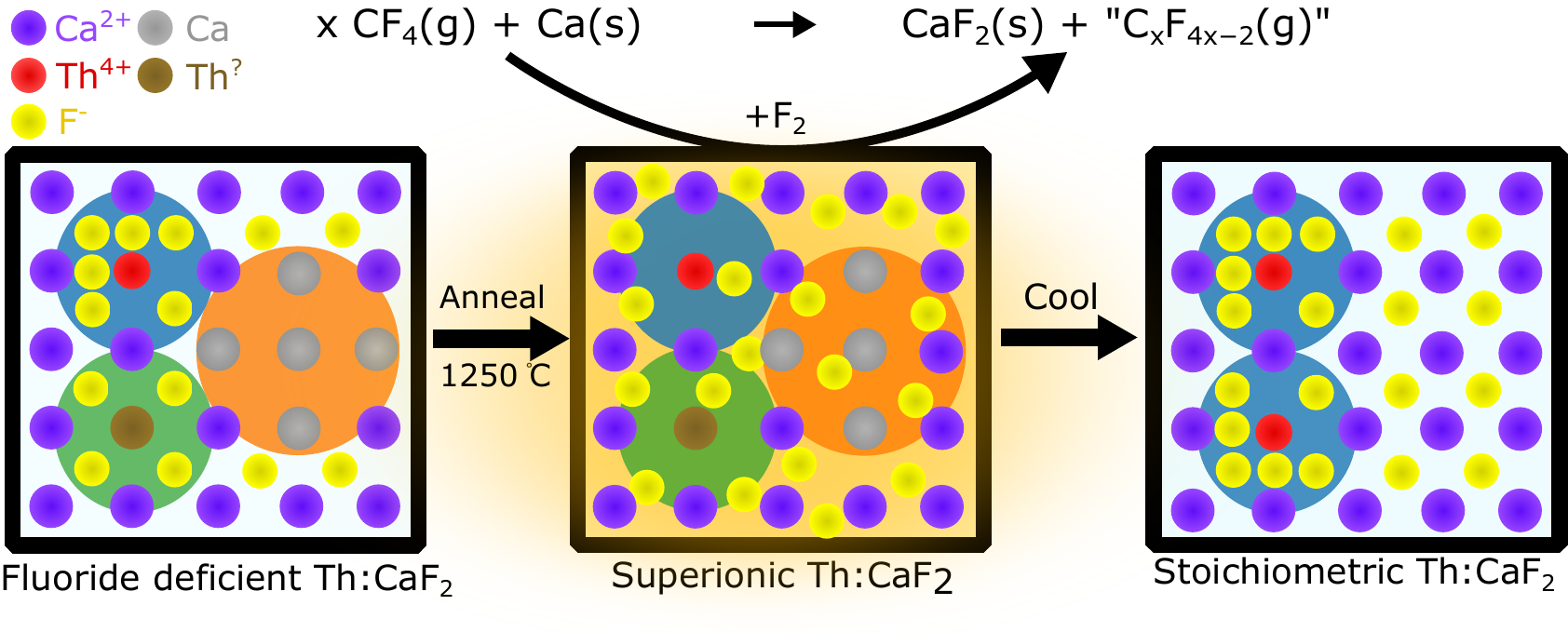}
    \caption{Schematic representation of the fluorination cycle and the corresponding change in local crystal structure. The fluoride deficient, imperfect crystal shows Ca metallic nanoparticles (orange region, right), Th doping with unknown surrounding and charge state (green region, bottom left), and Th stoichiometrically surrounded by fluoride ions (blue region, top left)~\cite{dessovic2014229thorium}. The associated absorption peaks are identified in the text and measured in \autoref{fig:229absorption}. Adding \ch{F} atoms at superionic temperatures distributes the anions quickly through the crystal due to the increased mobility~\cite{gillan1986collective}. After slowly cooling down, the crystal reaches a stoichiometric composition with the dopant in the ideal surrounding. The increase in blue absorption and decrease in all others is shown in \autoref{fig:absorbchange}.}
    \label{fig:cycle}
\end{figure*}

The process of growing \ch{CaF2} in a vacuum environment leads to a fluoride ion (\ch{F-}) deficit due to thermal dissociation of \ch{CaF2} and its reaction with the residual water in the system~\cite{recker1971zuchtung,bollmann1980incorporation}. This sequence leads to the formation of a non-stoichiometric or fluoride-deficient crystal. To counterbalance the loss in fluoride ions, the crystal tends to generate metallic Ca nanoparticles~\cite{beeks2022nuclear}, which possess the capacity to absorb and scatter light~\cite{rix2011radiation}, especially within the VUV and optical range. Non-stoichiometry in all ionic crystals leads to changes in the configuration: either change of charge state of cations or formation of metal colloids~\cite{hughes1979metal}. Although this variation in configuration has been thoroughly examined in oxides like \ch{CeO2}~\cite{li2021review,barth2016perfectly}, it has not been extensively studied for fluorides and there is a lack of common terminology, understanding and means of controlling the compositions~\cite{cirillo1987fluorine,dubois2011stabilization,wang2005synthesis,antonyak2017charge,Angervaks:18}. 

Due to the presence of a radioactive element during growth, radiolysis causes enhanced dissociation of \ch{CaF2}~\cite{beeks2023growth,schmedt2012occurrence}. The strong dissociation creates a very non-stoichiometric composition. The change in composition leads to a change of dopant configuration and absorption profile, as was first observed in the work of Cirillo et al. (1987)~\cite{cirillo1987fluorine}. Here, adding \ch{F2} to \ch{Eu^2+}:\ch{CaF2} changed the charge state of Eu and thus the absorption profile of the crystal. A lack of fluoride in the crystal is compensated for by reducing the calcium to neutral, and the dopant to a lower charge state. By adding fluoride the charge state of the dopant can be increased and the calcium oxidized. 

Radioactivity not only affects growth: (doped) \ch{CaF2} subjected to radiation from radioactive decay after growth shows complex behavior centered around fluoride motion in the lattice~\cite{celinski2016trace}. The \ch{^{229}Th}:\ch{CaF2} crystals therefore are a unique and complex system for which we give a concise description, following the literature~\cite{hayes2012defects,catlow1980irradiation,rix2011radiation,rodnyi2020physical,beeks2022nuclear}.

The presence of a radioactive element in the \ch{CaF2} matrix is a source of excitation of the lattice, which produces defects and aggregates which lie at the core of understanding the dynamic processes. Mainly the motion of fluorine determines the produced defects and their respective absorption and emission bands: fluoride moves through the crystal expending much less energy (0.6-\SI{1.5}{\eV}) than the calcium ions and the formation energy of an anion Frenkel pair (\ch{F^-} interstitial and vacancy) is \SI{2.7}{eV} as compared to \SI{6}{\eV} to form a cation Frenkel pair. Therefore the main defects in \ch{CaF2} are related to fluoride: F, H and V\textsubscript{k} centers. The F center is an electron trapped at the location of a fluoride vacancy, which is similar to an electron in a box. Therefore the F center has rich optical absorption and emission bands. Th H center is a fluoride interstitial, which shares a trapped hole with a lattice fluoride thereby creating an \ch{F2^-} dimer with high mobility. The V\textsubscript{k} center is a trapped hole shared between two lattice fluorides, thereby also creating an \ch{F2^-} dimer. 

The energy deposited by \ch{^{229}Th} and its daughters through $\alpha$ and $\beta$ decay is in the range of \SI{100}{\kilo\eV} to \SI{8.5}{\mega\eV} per decay. Such a high energy excitation will produce core shell holes and highly excited electrons. These will decay through various mechanisms such as Auger electrons, x-ray emission and plasmons to electron hole pairs (12.2 eV formation energy),  and self-trapped-excitons (STE, formation energy 11.18 eV, absorption bands at 282 and \SI{482}{\nm}, emission bands from 200 to \SI{500}{\nm}). The electron hole pair can annihilate under photon emission, non-radiatively create a separate F and H center in the lattice (imperfect damaged crystal) or decay to a STE. The STE initially constitutes of a self-trapped-hole (V\textsubscript{k} center) and a captured electron, but quickly decays to an F and H pair. The F-H pair again either annihilates and emits a photon, or decays non-radiatively and leaves a separate F and H center in the crystal.  

The F center (absorption at \SI{378}{\nm}, emission at \SI{585}{\nm}) agglomerates into higher order M (2 F), R (3 F) and N (4 F) centers and continues to form Ca metallic colloids (fluoride vacancy agglomeration in \ch{CaF2} is metallic Ca) as it is energetically favorable. The agglomeration shifts absorption and emission bands to lower energies. Most importantly the Ca colloids absorb in two bands, from 550 to \SI{960}{nm} and 160 to \SI{200}{\nm} depending on their size~\cite{Angervaks:18}. In thorium doped crystals these colloids absorb around \SI{150}{\nm}~\cite{beeks2022nuclear} due to the change in refractive index. 

The H center (absorption at \SI{310}{\nm}) can further collapse either to an impurity trapped hole (identified at \SI{295}{\nm} for Th) or two H centers form an interstitial dimer (two interstitial \ch{F^-} that share a hole,\ch{F2^-}). The dimer formation can also aggregate to form dislocation loops (as observed). Through hopping, the single (or higher order) F and H centers can again find one another and annihilate (evaporate) under photon emission, thereby again producing a perfect crystal. If a stoichiometric amount of Ca and \ch{F2} is present, this can be done through annealing at \SI{600}{\degreeCelsius}. If not, fluoride annealing needs to be applied.

Due to the constant internal irradiation of this crystal, the above processes are dynamic and the fluorides are in motion. Defects will accumulate until a steady state of growth and evaporation of Ca colloids and H center dimers is reached. The constant irradiation also provides a constant stream of Cherenkov photons (200 to \SI{122}{nm}). Irradiated non-stoichiometric \ch{CaF2} damages faster, as a larger amount of F and H centers is already present to compensate for non-stoichiometry, which can quickly aggregate upon irradiation to larger more absorbing defects. The damage steady state is reached quicker and annealing does not completely repair radiation damage. To create more transparent and radiation resistant \ch{CaF2}, fluorine needs to be added for a stoichiometric crystal.

Therefore we study the impact of the Th:\ch{CaF2} crystal composition on its VUV transmission and the electronic structure of the dopant sites. Due to the \ch{^{229}Th} radioactivity, radiolysis becomes the major cause of the fluorine (\ch{F2}) loss during the growth phase. This substantial \ch{F2} loss modifies the electronic structure of the \ch{CaF2} crystal, resulting in a consequential change in its absorption profile~\cite{beeks2023growth}. The large resulting VUV absorption would make such a \ch{^{229}Th}:\ch{CaF2} unsuitable for a nuclear optical clock.

We follow~\cite{cirillo1987fluorine} and developed a novel and safer experimental method that does not require toxic \ch{F2} gas to add fluoride ions to already grown, single-crystalline, Th doped \ch{CaF2} and still dramatically improve the transmission profile.

We use an induction heated carbon crucible to anneal Th:CaF\textsubscript{2} at above its superionic temperature, but below its melting temperature, in a carbon tetrafluoride (\ch{CF4}) atmosphere. The crystal is placed in the center of an induction coil in a carbon crucible, after which the system is evacuated to approximately \SI{1e-6}{\milli\bar}. The chamber is then filled with \SI{1.1}{\bar} \ch{CF4}. In the heating step, the carbon crucible is heated using the induction coil (\SI{1300}{\degreeCelsius}, heating rate \SI{20}{\kelvin\per\min}) while the walls of the vacuum system are water-cooled, creating a steep temperature gradient. The temperature gradient causes the \ch{CF4} gas to be reactive only at the crystal surface and inert at the vacuum system walls, which significantly reduces safety concerns while ensuring the efficiency of the process. After annealing for 1 hour while holding the temperature, the system is cooled down (cooling rate \SI{1}{\kelvin\per\min}) and at room temperature the \ch{CF4} is replaced by \ch{N2}. A schematic representation of the process is shown in \autoref{fig:cycle}. 

Above the superionic transition temperature of \ch{CaF2} (\SI{1250}{\degreeCelsius
}~\cite{gillan1986collective}), but below the melting temperature, fluoride anions exhibit high mobility while calcium cations remain immobile. The mobility of fluoride ions ensures their uniform distribution throughout the bulk crystal.

\begin{figure}[h!]
    \centering
    \includegraphics[width=\linewidth]{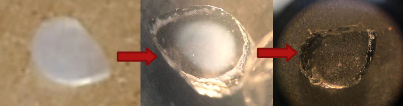}
    \caption{Annealing of a fluoride-deficient \textsuperscript{232}Th:CaF\textsubscript{2} crystal in two steps, each step lasting one hour. Ca colloids absorb and scatter heavily in the visible region~\cite{rix2011radiation}. It is clearly seen that after the first cycle the opaqueness recedes to the center, while after the second fluorination step the crystal has fully regained optical transmission.}
    \label{fig:fdefanneal}
\end{figure}

To produce a heavily fluoride deficient crystal, we first performed superionic annealing in vacuum instead of a \ch{CF4} atmosphere on a \ch{^{232}Th}:\ch{CaF2} crystal. Excessive fluorine loss from the crystal during vacuum superionic annealing for 24 hours has turned an initially transparent crystal into a cloudy and opaque one (see \autoref{fig:fdefanneal}, left). Annealing in an argon atmosphere was also performed, producing a similar but less deficient crystal. The formation of calcium metal colloids can explain the opacity \cite{rix2011radiation}. In the following phase, which consists of \ch{CF4} annealing in two cycles of one hour each, the crystal regains its visible transparency (see \autoref{fig:fdefanneal}, middle and right).

\begin{figure}[h!]
    \centering
    \includegraphics[width=\linewidth]{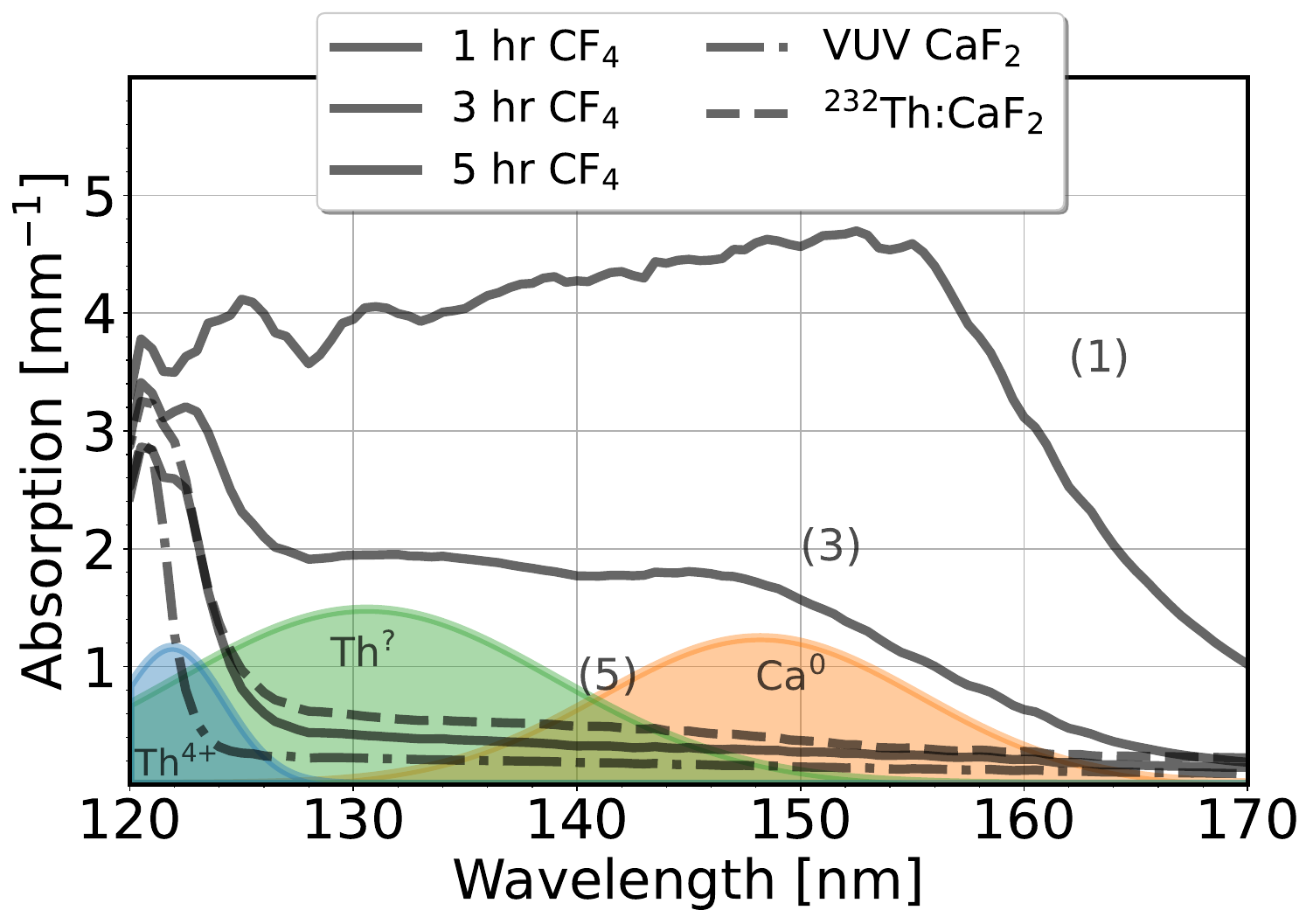}
    \caption{VUV absorption spectrum of a \SI{6e{18}}{\per\cubic\centi\metre} doped \ch{^{229}Th}:\ch{CaF2} crystal for different superionic fluoride annealing durations (numbers indicate annealing time). To compare, absorption of a \ch{^{232}Th}:\ch{CaF2} crystal with similar doping concentration without fluoride annealing is plotted and that of a VUV grade pure \ch{CaF2} sample. The three absorption line centers shown in \autoref{fig:cycle} are fitted with gaussian functions, while compensating for the undoped \ch{CaF2} absorption background and drawn in the figure with the same colors and indicated defect.}
    \label{fig:229absorption}
\end{figure}

Following the fluorination treatment, we observed a significant improvement in the VUV transparency of the \ch{^{229}Th}:\ch{CaF2} crystals, as shown in \autoref{fig:229absorption}. Note that this crystal was fully opaque around 8\,eV (150\,nm) and hence unusable for nuclear laser spectroscopy directly after growth. After several cycles, the absorption was lower as compared to the non-radioactive \ch{^{232}Th} doped crystal, indicating \ch{CF4} could improve its absorption profile as well (as was done in~\cite{gong2024structures}). When comparing this absorption spectrum to pure \ch{CaF2}, three absorption centers can be seen to appear and disappear. The absorption spectrum was recorded using a McPherson 204/302 VUV spectrometer, as described in the work of Beeks et al.~\cite{beeks2022nuclear}. As verified by gamma spectroscopy, the annealing process did not cause any quantifiable loss of radioactivity, indicating no noticeable reduction of the \textsuperscript{229}Th concentration. 

\begin{figure}[h!]
    \centering
    \includegraphics[width=0.9\linewidth]{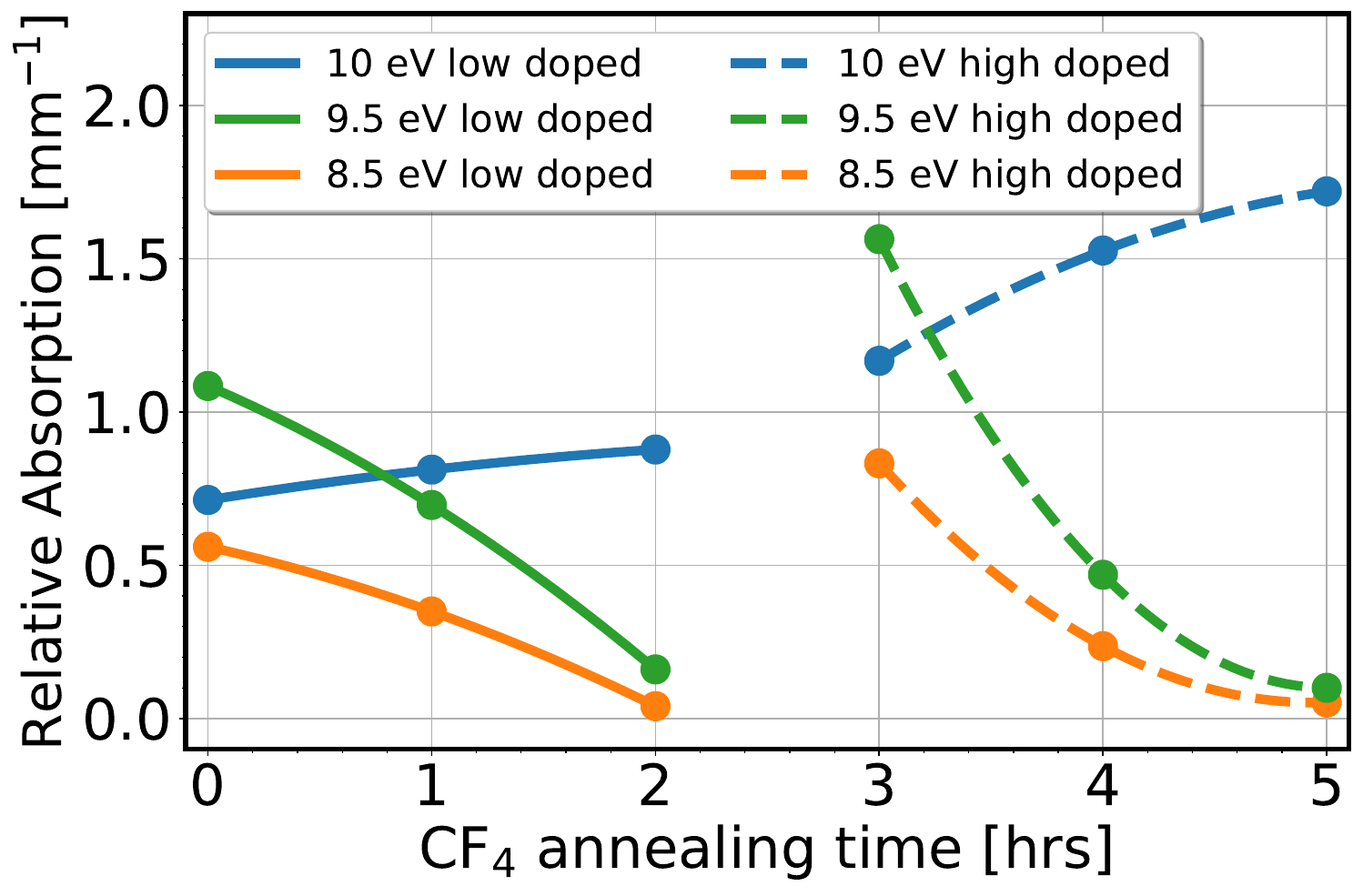}
    \caption{Intensities of the three absorption lines relative to undoped \ch{CaF2} identified in \autoref{fig:cycle} and \autoref{fig:229absorption} at \SI{121.9}{\nm}, \SI{130.5}{\nm} and \SI{148.2}{\nm} or approximately \SI{10}{\eV}, \SI{9.5}{\eV} and \SI{8.5}{\eV} is plotted as a function of superionic fluoride transfer time for two \ch{^{229}Th}:\ch{CaF2} crystals (\SI{1e{18}} and \SI{6e{18}}{\per\cubic\centi\metre}). Data points are connected with splines to lead the eye.}
    \label{fig:absorbchange}
\end{figure}

To quantitatively describe the VUV absorption profile over the time of fluorination, we identify three absorption lines (assuming gaussian profile) in \autoref{fig:229absorption}.
We attribute the \SI{122}{\nm} absorption to the \ch{Th^{4+}} charge transfer state~\cite{Nickerson2020color,nickerson2021driven}, the \SI{130}{\nm} absorption to \ch{Th} ions in a different surrounding, and the broad \SI{150}{\nm} absorption to \ch{Ca} metallic colloids~\cite{rix2011radiation,ryskin2017stabilization}. The \SI{130}{\nm} absorption is likely caused by a change in the surrounding of the Th ion: either the change of charge state of the Th atom to neutral/1+/2+/3+ due to the fluoride deficiency, reaction of the non-stoichiometric \ch{CaF2} with oxygen in the air thereby replacing the \ch{F} atoms surrounding \ch{Th} by \ch{O} atoms or a change in charge compensation mechanism to for example a \ch{Ca} vacancy~\cite{gong2024structures,pimon2022ab}. Investigations of the \SI{130}{\nm} absorption line are ongoing. In \autoref{fig:absorbchange} it can be seen that in low and high-doped crystals the \SI{122}{\nm} absorption increases with cycle number, indicating an increase in \ch{Th^{4+}} with fully charge-compensated surrounding. The other two absorption lines assigned to Ca colloids and Th in a different surrounding both diminish. Effectively the electronic structure of the Th dopants is manipulated by adding fluoride ions to the deficient crystal.

After fluorination of the \ch{^{229}Th}:\ch{CaF2}, no change in color is observed in the crystal over the course of a year as opposed to the original fluoride deficient crystals~\cite{beeks2023growth}. The VUV transmission decreased from 50\% to 35\% over the course of 1 year, as opposed to complete VUV opacity in 3 days for non-fluorinated crystals. The original transmission could be completely regained through superionic fluoride transfer. After a year, no other changes in the radioactive crystals are observed such as cracking, thus it is concluded the radiation hardness significantly increased after fluorination as predicted.

Alternative methods of fluorination of low-doped \ch{^{229}Th}:\ch{CaF2} crystals were also tested. Highly fluoride deficient crystals (opaque) were treated using 3 different methods. 

The first method was annealing in an \ch{F2} atmosphere. A Mg sample holder was passivated for two days at \SI{600}{\degreeCelsius} (heating rate: \SI{4}{\kelvin\per\hour}, cooling rate:  \SI{1}{\kelvin\per\minute}) in an \ch{F2} flow (20\,\% \ch{F2} in \ch{N2} atmosphere, \SI{5}{\cubic\cm\per\minute}). The \ch{^{229}Th}:\ch{CaF2} crystal was placed in the sample holder and was fluorinated for five days at \SI{600}{\degreeCelsius} (heating rate: \SI{4}{\kelvin\per\hour}, cooling rate: \SI{1}{\kelvin\per\minute}) in an \ch{F2} flow (20\,\% \ch{F2} in \ch{N2} atmosphere, \SI{5}{\cubic\cm\per\minute}). 

The second method was treatment using an \ch{NF3} plasma. In this method a Ni sample holder was passivated in a fluorine plasma at room temperature for one hour. The \ch{^{229}Th}:\ch{CaF2} crystal was placed in the sample holder and was fluorinated in a fluorine plasma at room temperature for three hours. \ch{NF3} was used as feeding gas with a flow rate of \SI{30}{\cubic\cm\per\minute}. 

The third method was treatment in a \ch{F2} filled autoclave. In this method a Ni sample holder was passivated for one day at \SI{400}{\degreeCelsius} and approximately \SI{400}{\bar} (heating rate: \SI{100}{\kelvin\per\hour}, cooling rate: \SI{4}{\kelvin\per\minute}, 20\% \ch{F2} in Ar atmosphere). The \ch{^{229}Th}:\ch{CaF2} crystal was placed in the Ni sample holder and was high-pressure fluorinated at \SI{400}{\degreeCelsius} and approximately \SI{400}{\bar} for five days (heating rate: \SI{100}{\kelvin\per\hour}, cooling rate: \SI{4}{\kelvin\per\minute}, 50\% \ch{F2} in Ar atmosphere). 

\begin{figure}[h!]
    \centering
    \includegraphics[width=\linewidth]{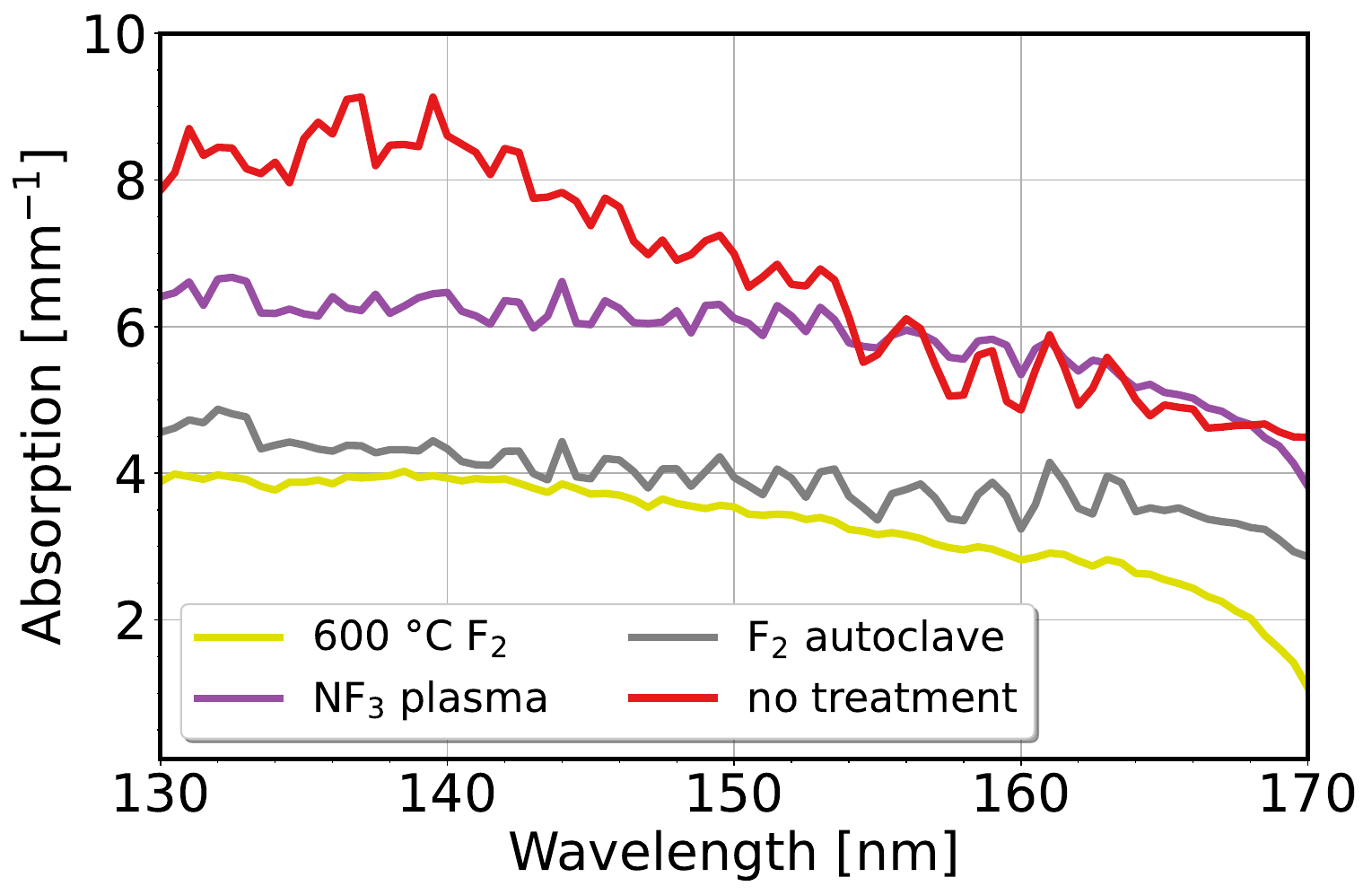}
    \caption{Absorption spectra of different segments of a \SI{1e{-17}}{\per\cubic\centi\metre} doped \ch{^{229}Th}:\ch{CaF2} crystal that was grown using \SI{1}{\mega\becquerel} of activity for different fluoride treatments. No reliable data could be obtained below \SI{130}{\nm} due to the high absorption.}
    \label{fig:treatments}
\end{figure}

The resulting crystals still displayed high absorption as can be seen in \autoref{fig:treatments}. The 5 days of \ch{F2} annealing at \SI{600}{\degreeCelsius} displayed results similar to 1 hour of \ch{CF4} annealing at \SI{1250}{\degreeCelsius}.

Compared to alternative treatments, only superionic annealing under \ch{CF4} atmosphere demonstrated a significant improvement in the VUV transmission (compare \autoref{fig:229absorption} with \autoref{fig:treatments}). Consequently, it can be concluded that the superionic state of \ch{CaF2} allows for efficient and rapid homogeneous distribution of the acquired fluoride ions. This contrasts with other methods where the solid phase of the crystal prevents fluoride ion diffusion through the crystal.

Many facts have been gathered on \ch{^{229}Th}:\ch{CaF2}: The observed absorption centers at 122, 130 and \SI{150}{\nm}, their change with fluoride content and the known fluorescence of \ch{^{229}Th}:\ch{CaF2} at \SI{184}{\nm}~\cite{kraemer2023observation} (and at 168, 230, 238, 250 and \SI{295}{\nm}~\cite{beeks2022nuclear}). Aside from the \SI{150}{\nm} absorption which has been identified as \ch{Ca} nanoparticles~\cite{rix2011radiation}, we can speculate on the origin of the other absorption and fluorescence bands. We will compare thorium with chemical analogues such as cerium and hafnium. 

The cerium analogue we will use to compare spectroscopic data in \ch{CaF2}, as its 4+ and more common 3+ states have similar electronic configuration. Due to relativistic effects, the differences between Ce and Th might be large. The hafnium analogue we will use for the charge state thorium assumes, as neutral hafnium has a similar valence electron configuration ($5d^26s^2$ vs. $6d^27s^2$) and chemically behaves similar to thorium. Just as thorium, hafnium almost exclusively takes a 4+ charge state, as opposed to cerium. 

It is known that the rare earths in \ch{CaF2} reduce upon X-ray irradiation~\cite{kiss1965dynamics}.\ch{Hf^{4+}} doped in \ch{YPO4} can be reduced to \ch{Hf^{3+}} by X-ray irradiation~\cite{laguta2019electron} and it is known that the actinides reduce through self-irradiation in \ch{CaF2}~\cite{stacy1972effects}. We thus speculate that thorium undergoes the same process from 4+ to 3+. As F centers and electrons in the conduction band are constantly produced, the \ch{Th^4+} cations will attract them through the coulomb force more strongly compared to \ch{Ca^2+} cations. Through the change of absorption intensities with fluoride annealing we assert that the change in charge state is not only through irradiation but also stoichiometry. It can be argued that these are the same processes: Locally, the creation of F centers through irradiation creates non-stoichiometry thereby changing the charge state of the dopant. 

Comparing \ch{Th^{3+}} to \ch{Ce^{3+}} doped in \ch{CaF2}, we can find that \ch{Ce^{3+}} has several $4f^n$ to $4f^{n-1}5d$ excitations around \SI{180}{\nm}~\cite{van2002light,yamaga2004optical}. The bare \ch{Th^{3+}} ion has a strong $6d$ to $7p$ line at \SI{170}{nm}~\cite{KLINKENBERG1949774}. Fluorescence lines in the VUV are known to exist for 3+ heavy rare earth ions in \ch{CaF2}~\cite{van2002heavy}. We speculate that the fluoride deficient \ch{CaF2} contains some \ch{Th^{3+}}, that is optically active through absorption at \SI{130}{nm} and emission at 168 and \SI{180}{nm}. Through addition of fluoride these disappear and only the \ch{Th^{4+}} absorption at \SI{120}{\nm} remains. This charge transfer state was predicted~\cite{Nickerson2020color,nickerson2021driven} and would constitute creating an electron hole pair on the thorium dopant and a neighboring fluoride. We speculate that this pair decays to a defect stabilized V\textsubscript{k} plus electron center~\cite{beaumont1970investigation}, which emits at \SI{295}{\nm}~\cite{beeks2022nuclear}.

This report demonstrates the enhancement of optical transmission in ionic crystals through superionic fluoride transfer. This superionic state substantially decreases the treatment duration. Furthermore, our findings suggest the possibility of controlling the dopant surrounding in \ch{CaF2} by the addition or removal of \ch{F-}, although the removal process carries a potential risk of creating \ch{Ca} metallic nanoparticles. We have developed a simple and safe method of fluorine manipulation in fluoride ionic crystals. Utilizing this method, we are able to fabricate highly transparent, heavily doped \ch{^{229}Th}:\ch{CaF2} crystals for a solid-state nuclear clock. The manipulation of the electronic structure brings the potential for advancements in optics, scintillator and laser crystal development by optimizing light absorption and emission through dopant and cation surrounding control (e.g., \ch{Eu}:\ch{CaF2}).
\subsection{Acknowledgements}
\begin{acknowledgments}
This work is part of the thorium nuclear clock project that has received funding from the European Research Council (ERC) under the European Union's Horizon 2020 research and innovation programme (Grant Agreement No. 856415). The research was supported by the Austrian Science Fund (FWF) Projects: I5971 (REThorIC) and P 33627 (NQRclock).
\end{acknowledgments}
\bibliography{ms.bib}

\end{document}